# Nanostructured potential well/barrier engineering for realizing unprecedentedly large thermoelectric power factors

Neophytos Neophytou[1,*], Samuel Foster[1], Vassilios Vargiamidis[1], Giovanni Pennelli[2], and Dario Narducci[3]

[1]School of Engineering, University of Warwick, Coventry, CV4 7AL, UK
[2]Department of Information Engineering, University of Pisa, Pisa, I-56122, Italy
[3]Dept. of Materials Science, Univ. of Milano Bicocca, Milano, I-20125, Italy
[*]e-mail: N.Neophytou@warwick.ac.uk

## Abstract

This work describes through semiclassical Boltzmann transport theory and simulation a novel nanostructured material design that can lead to unprecedentedly high thermoelectric power factors, with improvements of more than an order of magnitude compared to optimal bulk material power factors. The design is based on a specific grain/grain-boundary (potential well/barrier) engineering such that: i) carrier energy filtering is achieved using potential barriers, combined with ii) higher than usual doping operating conditions such that high carrier velocities and mean-free-paths are utilized, iii) minimal carrier energy relaxation after passing over the barriers to propagate the high Seebeck coefficient of the barriers into the potential wells, and importantly, iv) the formation of an intermediate dopant-free (depleted) region. The design consists thus of a 'three-region geometry', in which the high doping resides in the center/core of the potential well, with a dopant-depleted region separating the doped region from the potential barriers. It is shown that the filtering barriers are optimal when they mitigate the reduction in conductivity they introduce, and this can be done primarily when they are 'clean' from dopants during the process of filtering. The potential wells, on the other hand, are optimal when they mitigate the reduced Seebeck they introduce by: i) not allowing carrier energy relaxation, and importantly ii) by mitigating the reduction in mobility that the high concentration of dopant impurities cause. It is shown that dopant segregation, with 'clean' dopant-depletion regions around the potential barriers, serves this key purpose of improved mobility towards the phonon-limited mobility levels in the wells. Using quantum transport simulations based on the non-equilibrium Green's function method (NEGF) as well as semi-classical Monte Carlo simulations we also verify the important ingredients and validate this 'clean-filtering' design.

I. Introduction

Thermoelectric (TE) materials have made dramatic progress over the last several years. The thermoelectric figure of merit *ZT*, which quantifies the ability of a material to convert heat into electricity, has more than doubled compared to traditional values of *ZT*~1, reaching values above *ZT*~2 in several instances across materials and temperature ranges [1, 2, 3, 4, 5, 6, 7, 8, 9, 10, 11, 12]. The figure of merit is determined by $ZT=\sigma S^2 T/(\kappa_e+\kappa_l)$, where $\sigma$ is the electrical conductivity, *S* is the Seebeck coefficient, *T* is the absolute temperature, and $\kappa_e$ and $\kappa_l$ are the electronic and lattice parts of the thermal conductivity, respectively. The recent improvements in *ZT* are mostly attributed to drastic reductions of the lattice thermal conductivity in nanostructured materials and nanocomposites which has reached amorphous limit values at $\kappa_l$ = 1-2 W/mK and below [1, 3, 13, 14], as well as complex phonon dynamics materials [4]. With such low thermal conductivities, however, any further benefits to *ZT* must be achieved through the improvement of the thermoelectric power factor (PF) $\sigma S^2$, for which no noticeable progress has so far been achieved.

The lack of progress in the power factor improvement is attributed to the adverse interdependence of the electrical conductivity and Seebeck coefficient via the carrier density, which proves very difficult to overcome. The most commonly explored direction in improving the power factor is the 'conventional' energy filtering approach in nanocomposites and superlattices, in which built-in potential barriers block the cold low energy carriers, while allow the hot high energy carriers to flow [15, 16, 17, 18, 19, 20, 21, 22, 23, 24, 25, 26, 27, 28, 29, 30]. This energetic preference increases the Seebeck coefficient. On the other hand, this approach has still not been widely applied because the conductivity is also reduced in the presence of potential barriers. Current research efforts in improving the power factor have thus diverted into many other directions, including: i) taking advantage of the density of states in low-dimensional materials through quantum confinement [31], or in bulk materials that include low-dimensional 'like' features [32, 33], ii) bandstructure engineering and band-convergence strategies [32, 33, 34, 35, 36], iii) modulation doping and gating [37, 38, 39, 40, 41, 42, 43, 44, 45], iv) introducing resonances in the density of states [46], and even more recently v) concepts that take advantage of the Soret effect in hybrid porous/electrolyte materials [47]. These approaches target improvements either in the Seebeck coefficient or the electrical conductivity, with the hope that the other quantity will not be degraded significantly, and sometimes they report moderate power factor improvements. However, no significant developments that lead to meaningful improvements, wider applicability, or generalization to many materials have been achieved by these methods either.

Recent efforts by us and others, however, both theoretical and experimental, have revisited the energy filtering concept, and targeted designs that provide simultaneous improvements in both the electrical conductivity and the Seebeck coefficient in order to largely improve the power factor $\sigma S^2$. Experimental works have indeed verified that it is possible to achieve very high power factors in nanostructured Si-based materials after undergoing specific treatment [48, 49, 50, 51, 52]. Measured data for PF improvements of over 5× compared to bulk values, were adequately explained using Boltzmann transport theory [48, 49]. In those cases the grain boundaries of the nanocrystalline material serve as potential barriers resulting in energy filtering and high Seebeck coefficients. On the other hand, the carrier mean-free-path for scattering (and mobility) in the grains was improved due to the fact that dopants were primarily placed in the middle regions of the grains, rather than uniformly spread in the material – see later on Fig. 3a. The regions around the grain boundary potential barriers were depleted of dopants and allowed higher local carrier mobility and higher mean-free-paths, which resulted in higher grain conductivities and overall material conductivities [48]. Thus, compared to other approaches, recent evidence suggests that energy filtering could be engineered in such a way as to lead to large PF improvements.

In light of the strong evidence of such exceptionally high power factors demonstrated, as well as the recent surge in efforts to use energy filtering and design the grain/grain-boundary system efficiently in a variety of materials [28, 29, 30, 53], in this work, we re-examine theoretically energy filtering under degenerate conditions and dopant segregation. Using Boltzmann transport theory, we provide an in-depth investigation of a generalized grain/grain-boundary design concept, and explore the ultimate upper limit that can potentially be achieved under realistic conditions. The design is based on a number of concepts/'ingredients', whose contributions are examined individually, and then all combined until the final upper limit is reached. To further examine the validity of those 'ingredients' we also employ more sophisticated quantum transport simulations and Monte Carlo simulations. We show that power factors even up to the exceptionally high value of $PF \sim 50$ mW/mK$^2$ can be achieved once a nanostructure is properly designed.

The paper is organized as follows: Section II describes the main features of the proposed design, based on what we refer to as '3-region, clean-filtering'. Section III describes and validates the semiclassical transport model against p-type Si (the semiconductor material for which the largest improvements were experimentally achieved, but without loss in generality for the material choice). We then explore the power factor improvements that the individual design 'ingredients' lead to as they are added to the design one by one. Section IV increases the values of the design ingredients to very high, but still realistic levels, to provide an upper limit estimate for the power factor. Section V then discusses the validity of the assumptions/'ingredients' using the Non-Equilibrium Green's Function (NEGF) quantum transport method and Monte Carlo simulations. Finally, Section VI concludes the work.

## II. Approach

To model TE transport in the structure we examine, we start with the Boltzmann transport equation (BTE) formalism, and calibrate our model to match the mobility of p-type bulk Si so that we remain within realistic exploration boundaries. Within the BTE, the conductivity and Seebeck coefficient can be extracted as [54, 55]:

$$\sigma = q_0^2 \int_{E_0}^{\infty} dE \left( -\frac{\partial f_0}{\partial E} \right) \Xi(E) \equiv \int_{E_0}^{\infty} \tilde{\sigma}(E) dE, \tag{1a}$$

$$S = \frac{q_0 k_B}{\sigma} \int_{E_0}^{\infty} dE \left( -\frac{\partial f_0}{\partial E} \right) \Xi(E) \left( \frac{E - E_F}{k_B T} \right) \equiv \frac{1}{\sigma} \int_{E_0}^{\infty} \tilde{S}(E) dE, \tag{1b}$$

where $q_0$ is the electron charge and $k_B$ is the Boltzmann constant. Note that for easiness we define $\tilde{\sigma}(E)$ and $\tilde{S}(E)$ (here and throughout the text), as quantities which when integrated over energy give the conductivity and Seebeck coefficients $\sigma$ and $S$ (thus, they do not have the units of conductivity and Seebeck coefficient themselves). The transport distribution function $\Xi(E)$ in Eq. 1a is defined as:

$$\Xi(E) = \tau(E)\upsilon(E)^2 g(E), \tag{2a}$$

$$\tau(E) = \frac{\lambda_0 (E/k_B T)^r}{\upsilon(E)}, \tag{2b}$$

where $\tau(E)$ is the momentum relaxation time, $\upsilon(E)$ is the bandstructure velocity, $g(E)$ is the density of states, and $\lambda_0$ is the mean-free-path (MFP) for scattering. In Eq. (2b) the energy dependence of the MFP is introduced with a characteristic exponent $r$ that defines a specific scattering mechanism. In the case of phonon scattering the MFP is energy independent, $r = 0$ (for 3D channels), and consequently the scattering rate is proportional to the density of states. Note that Eq. 2b originates directly from the definition of the mean-free-path, which is defined as the average time between momentum relaxing scattering events $\tau(E)$ × the carrier bandstructure velocity $\upsilon(E)$ as $\lambda(E) = \upsilon(E)\tau(E)$. The mean-free-path then can be expressed by a constant, $\lambda_0$, times an energy dependent term [56, 57, 58]. The constant is adjusted to fit experimental low-field mobility data at low concentrations to mimic the transport properties of a specific material. In 3D, under isotropic scattering conditions (as for acoustic/optical phonons), the scattering rate is proportional to the density of final states which has an $E^{1/2}$ trend. With the velocity following an $E^{1/2}$ trend as well, then $\lambda(E) \propto E^{1/2}E^{-1/2} = E^0$, which results in the MFP being energy independent.

For ionized impurity scattering (IIS), different expressions for the scattering rate and for the screening length apply at different doping concentrations [56]. Here we use the Brooks-Herring model with screening, which is satisfactory for doping concentrations up to $10^{18}$ cm$^{-3}$ for mapping to the Si p-type mobility [55]. Above that concentration (which is more relevant to our design), we merge to the strongly screened transition rate as described in Ref. [48], in an effort to match as close as possible the measured mobility of p-type Si [59, 60, 61]. Figure 1a shows mobility calculations of p-type Si for the phonon-limited case (red line), and the phonon plus ionized impurity scattering limited case (blue line). By using $\lambda_0 = 7.4$ nm for phonons, the desired bulk low-density mobility for p-type Si at 300 K is achieved ($\mu = 450$ cm$^2$/V-s). Our results agree particularly well at the carrier concentrations of interest, around $p = 5\times10^{19}$ cm$^{-3}$. The difference between the phonon-limited mobility and the phonon plus ionized impurity scattering mobility, which can be up to an order of magnitude, is a central aspect of our design as will be explained below.

With regards to thermoelectric performance, Fig. 1b, 1c, and 1d show comparison of the phonon-limited to the phonon plus IIS conductivity, Seebeck coefficient, and thermoelectric PF, respectively for the p-type bulk Si where our model was calibrated on. The power factor $\sigma S^2$ peaks at high carrier concentrations around $10^{19}$-$10^{20}$/cm$^3$, and in most cases this is achieved by doping, which introduces strong IIS and severely limits the carrier mobility and electronic conductivity. The power factor would be at least a factor of >2× higher in the absence of IIS, which is the motivation behind modulation doping and gating methods [37, 38, 39, 40, 41, 42, 43, 44, 45]. To date, however, in such studies, the improvement over doped materials was only modest, in the best case the power factor values were similar to those of the doped materials. We show below, however, how this can be utilized in a very efficient way.

Energy filtering through the introduction of potential barriers, is the most commonly used approach to impede the flow of low energy carriers, and increase the Seebeck coefficient. As indicated in the schematic of Fig. 2a, electrons transport in the material through alternating potential wells and potential barriers. High energy electrons gain energy to jump over the barriers and make it to the next well region. Once the electrons pass over the barrier and enter the potential well, there they tend to lose energy, usually through the emission of optical phonons (Fig. 2a), and relax to the Fermi level of the well within a few optical phonon emission mean-free-paths, $\lambda_E$. Low energy electrons are blocked. In the sections below, we explain the model for electronic transport in the nanostructures we consider, under different geometrical conditions, which turn out to influence the power factor. We describe the equations that describe electronic transport in the potential wells and afterwards in the potential barriers, determining energy filtering.

<u>Well region - Filtering and electron flow in the potential well:</u> When $\lambda_E$ is small ($\lambda_E \ll L_W$, where $L_W$ is the length of the well), transport in the barrier/well regions is essentially independent of each other (Fig. 2b). In this case the total Seebeck coefficient and conductivity can be thought of as a simple combination of the individual quantities in each region, weighted by the size (volume fraction) of each region ('W'-well, 'B'-barrier) as [37]:

$$\frac{v_{tot}}{\sigma_{tot}} = \frac{v_W}{\sigma_W} + \frac{v_B}{\sigma_B} \tag{3a}$$

$$S_{tot} = \frac{S_W v_W + S_B v_B}{v_W + v_B} \tag{3b}$$

where in 3D bulk materials the weighting factor $v_i$ is the volume of each region. The derivation of a generalized composite Seebeck coefficient equation as above can be found in the Appendix. The 'local' TE coefficients of the wells in this case are computed by simply considering the bulk value of these regions as:

$$\sigma_W = \int_0^\infty \tilde{\sigma}_W(E) dE, \tag{4a}$$

$$S_W = \frac{1}{\sigma_W} \int_0^\infty \tilde{S}_W(E) dE, \tag{4b}$$

and those of the barriers by:

$$\sigma_B = \int_{V_B}^\infty \tilde{\sigma}_B(E) dE, \tag{4c}$$

$$S_B = \frac{1}{\sigma_B} \int_{V_B}^\infty \tilde{S}_B(E) dE. \tag{4d}$$

In this case, PF improvements cannot be easily achieved, as the improvement in the Seebeck coefficient is limited to the volume that the barriers occupy, and in those regions the resistance is significantly increased (exponential drop in density).

<u>Well region - Energy non-relaxing case:</u> In the opposite scenario with respect to carrier energy relaxation, i.e. $\lambda_E > L_W$, electrons can flow over the barriers without (fully) relaxing their energy when they enter the wells (Fig. 2c). In this scenario, only carriers with energies above the barrier height $V_B$ are contributing to transport. The Seebeck coefficient is benefitted in this case as it is by definition the average energy of the current flow <E> as [26]:

$$S = \left(\frac{\langle E \rangle - E_F}{q_0 T}\right). \tag{5a}$$

The fact that the flow is high in energy throughout the channel (barriers and wells) essentially results in propagating the high Seebeck of the barrier into the well. In this energy non-relaxing case the TE coefficients in the well region can then be computed by only considering transport above $V_B$ as:

$$\sigma_W = \int_{V_B}^{\infty} \tilde{\sigma}_W(E) dE \tag{6a}$$

$$S_W = \frac{1}{\sigma_W} \int_{V_B}^{\infty} \tilde{S}_W(E) dE. \tag{6b}$$

Here, the only thing that changes from the treatment of each region independently, is that the integral for the conductivity of the well begins from $V_B$, rather than from the band edge of the well $E_C = 0$ eV.

In typical semiconductors, $\lambda_E$ is in the order of several nanometers to a few tenths of nanometers. Therefore, nanostructures in which the wells are a few tenths of nanometers, operate between the two cases of Fig. 2b and 2c. Figure 2d a more realistic scenario of semi-relaxation of carriers. What is shown here is a quantum mechanical electronic transport simulation based on the Non-Equilibrium Green's Function method, including electron scattering with optical phonons in a superlattice-type structure. The dashed-blue line indicates the current flow (average of the yellow/green-flow colormap). The red-solid line indicates the average energy of the current flow <E(x)>, the quantity which when integrated and scaled over the channel provides the Seebeck coefficient. As observed, the current flows over the barriers by absorbing optical phonons, and afterwards relaxes down in energy in the wells by emitting optical phonons. Since $\lambda_E$ in these simulations is set to $\lambda_E$ = 15 nm and the length of the well to $L_W$ = 50 nm, the current energy is not completely relaxed in the well, instead, this corresponds to the semi-relaxed case.

<u>Barrier region - Momentum relaxing case:</u> Considering the design of the barrier regions now, the carriers can completely relax on the barriers (both their momentum and energy), in which case they are treated as independent regions from the wells. The right side of Fig. 2e shows the case where a carrier from the same state as before, 'relaxes' on the bandstructure of the barrier (blue line), then flows over the barrier and 'relaxes' again on the 'local' band $E(k)$ of the well. In this case, the barrier imposes a large resistance on the current flow, and damages the overall conductivity of the material. The TE coefficients for the barrier can then be computed as in Eqs. 4c-4d.

<u>Barrier region – Thermionic emission (momentum non-relaxing) case:</u> However, if the barrier is narrow, carriers can be thermionically (ballistically) emitted over the potential barrier without relaxing on the top of it. The left side of Fig. 2e shows the case of thermionic emission, where the carriers from a given bandstructure state flow ballistically over the barrier and end in the same $E(k)$ in the right side of the barrier (red arrow). We note that in reality for narrow barriers the 'local' bandstructure could be different compared to the bandstructures of the constituent materials, and quantum mechanical reflections will appear because of a degree of well/barrier state mismatch, which would add to interface resistance. These can be mitigated somewhat by employing smoothened barrier edges and oblique potentials (as in Fig. 3) rather than sharp ones [23, 24], and we address this issue later on. The TE coefficients on the barrier region under thermionic emission conditions can then be computed at first order by allowing only the carriers from the wells with energies higher than the barrier heights $V_B$ over the barrier (ignoring contact resistance at this point) as:

$$\sigma_B = \int_{V_B}^{\infty} \tilde{\sigma}_W(E) dE \tag{7a}$$

$$S_B = \frac{1}{\sigma_B} \int_{V_B}^{\infty} \tilde{S}_W(E) dE. \tag{7b}$$

## III. Nanostructured grain/grain-boundary design for very high PFs

<u>Designing optimal attributes for the filtering barrier:</u> After describing the basic transport features and equations in SL structures, we begin our investigation by simulations of the optimized attributes for the potential barrier, and then for the potential well – targeting PF improvements. Figure 4, shows row-wise the TE coefficients electrical conductivity $\sigma$, Seebeck coefficient $S$, and power factor, respectively. Column-wise, it shows the change in these coefficients after each step of the design process we consider. We begin with the pristine channel, whose TE coefficients are shown by the black-dashed lines in all sub-figures for comparison, in which case the doping is uniformly distributed in the entire channel. The basis structure we consider has a well size $L_W = 30$ nm, a barrier length $L_B = 2$ nm, and barrier height $V_B = 0.15$ eV (geometrical features which experimentally showed large power factors [48, 49, 51]).

As a first step, in the data of the first column of Fig. 4 (Fig. 4a-4c), we introduce a potential barrier, and consider independent transport in the well and barrier regions (i.e. full energy relaxation in the wells), as in the schematic of Fig. 2b. The TE coefficients for this system are given by Eq. 4. For the case where the entire structure is considered doped, the results shown by the green lines, indicate strong reduction in $\sigma$, a slight increase in $S$ as expected, but finally a strong reduction in the power factor, at least for carrier concentrations up to $\sim 10^{21}$.

In a second step, we consider the possibility that the potential barrier is free of dopants. Note that although the potential wells need to be doped to achieve the required carrier density, the barriers do not. In fact, other than them being formed due to band-edge discontinuities between the two materials (well and barriers), they can also be created by junctions of highly-doped/intrinsic regions, as shown in Fig. 3c-3d. The rationale behind the undoped barrier regions originates from the much higher phonon-limited mobility and mean-free-paths for scattering compared to those of the ionized impurity-limited transport (as shown in Fig. 1a). Here, the TE coefficients for the barrier can be computed as in Eq. 4c-4d, but the scattering time is only determined by electron-phonon scattering. In the case of an undoped barrier (Fig. 4a-4c red lines), $\sigma$ is degraded much less, $S$ increases slightly, and the power factor experiences a slight increase. Thus, as the barrier formation increases the material's resistance, by keeping it 'clean' from dopants, at least the mean-free-paths are longer, and a portion of the conductivity is restored, mitigating the reduction in conductivity. This allows higher mobility carriers, limited locally by phonon scattering alone, rather than by the much stronger ionized impurity scattering (see Fig. 1a).

The third barrier attribute we investigate is the case where the carriers undergo thermionic emission above the barrier (blue lines), i.e. the case where the barrier size $L_B$ is much narrower compared to the momentum relaxation mean-free-path of the carriers. Under the assumption of thermionic emission, the carriers from the wells that are not filtered and flow over the barriers, as indicated in the left of **Fig. 2e,** although subject to the doping in the wells, they occupy high velocity states. In this case, similarly to the undoped barrier case, slight power factor improvements are observed (blue line in Fig. 4c). The transport details over the barrier, however, are different in the two cases. In the undoped barrier case the resistance is mitigated by the use of the higher mobility 'dopant-free' barrier region. In the thermionic case, the resistance is mitigated by the use of higher velocity carriers coming from the well, rather that the lower velocity carriers from the top of the barrier. *The key outcome here, is that 'clean' barrier regions or thermionic transport over them, can restore the conductivity reduction that the potential barrier causes (for reasonable $V_B$ values).* We now proceed with examining the optimal attributes of transport in the potential well.

Designing optimal attributes for the well – (a) avoid energy relaxation: We next investigate electronic transport in the well region. At this stage we consider that the well region is a uniformly, highly doped region. From here on we assume the undoped barrier model, where charge carriers relax on the barriers. This is the conservative worst case scenario compared to the thermionic emission assumption, but we also examine thermionic emission as an upper limit scenario later on as well. As illustrated earlier in Fig. 2d, the current in a large portion of the well adjacent to the barriers propagates at larger energies (before it relaxes at lower energies), indicating both larger Seebeck coefficient and larger carrier velocities (and conductivity). In fact, it is these regions that provide power factor improvements in superlattices and nanocomposites, because the high Seebeck coefficient of the barriers propagates into the wells, and high energies keep the conductivity still high. The optimal design is achieved when the carriers do not relax their energy into the well region after they overpass the barrier as in Fig. 2c. In practice, the barriers are placed at distances short enough to enforce less relaxation, but long enough to reduce the resistance introduced. Thus, an optimal compromise is achieved under semi-relaxation conditions. The advantage of energy semi-relaxation to the Seebeck coefficient is described by us and others in several works [27, 26, 62, 63, 64].

The TE coefficients in the limit of no-relaxation are shown by the red line in the second column of Fig. 4, in Fig. 4d-f. Essentially when the carriers flow at high energies in the wells, they raise the wells' low Seebeck coefficient to the high Seebeck values of the barriers (Fig. 4e), extending this high Seebeck in the entire channel. The downside is that only carriers with energies above $V_B$ are utilized, which would reduce the conductivity of the well and the overall channel conductivity. However, this is mitigated by the fact that the carriers of high energies are carriers of high mobility. The inset of Fig. 4d shows the mobility of p-type Si (dashed line), and the mobility of the carriers that travel above $V_B$ alone, which in this case is more than double (solid line). The conductivity is significantly lower compared to the pristine channel for lower densities (Fig. 4d), but as the Fermi level is raised to the $V_B$ level, the conductivity increases substantially. At those carrier densities the Seebeck coefficient (Fig. 4e) is still high, and therefore the overall power factor is improved compared to bulk by ~2× (Fig. 4f). This again illustrates the benefits of filtering at degenerate conditions [23, 48].

Thus, to summarize the design of the filtering well/barrier system up to this point, one seeks for material designs which: i) Allow for momentum non-relaxation on the top of the barrier, or 'clean' preferentially undoped barriers that restore/mitigate the conductivity reduction introduced by the potential barriers. ii) Allow for energy non-relaxation in the well, which raises the low Seebeck of the wells to the high values of the barriers. Those criteria impose restrictions in the design, shape, and importantly the 'cleanliness' of the barrier (from dopants for example), to achieve large momentum relaxation lengths, and potential well sizes comparable to $\lambda_E$ (or somewhat larger).

<u>Filtering well/barrier optimization – a novel concept consisting of three regions:</u> Moving forwards, a novel filtering design geometry is introduced, and in the following sections its performance is investigated. Although for the analysis we still employ the semi-classical Boltzmann transport formalism, later on quantum mechanical simulators and Monte Carlo simulators are also utilized to further validate some of the design 'ingredients'. A 2D top-down view of the proposed geometry is illustrated in Fig. 3a, where we now have rectangular domains of wells depicted by the blue colored regions and barriers depicted by the grey colored regions. The wells can represent highly doped regions, or regions where the band energy is in general lower (such that wells are formed).

The red colored regions in between the heavily doped regions and the barrier regions are part of the wells (as in a nanocomposite material, for example), although for those regions we consider that the doping is different; for the purposes of this analysis, these regions are undoped. A simplified schematic of the potential profile in a 1D cross section of the material is shown in Fig. 3b, with the middle, doped regions being lowered in energy, the barriers regions' energy residing higher, and the potential of the middle region connecting the two extremes. Here we do not explore the details of the formation of the band profile in this region, but our goal is to illustrate the design principle.

In practice, the oblique potential profile in the middle region can be, for example, a result of the n++/i junction that is formed, pushing most of the depletion region in the undoped, intrinsic part. Its formation will be dictated by self-consistent electrostatics and can be extracted by solving the Poisson's equation together with carrier statistics as shown in Fig. 3c and 3d. In Fig. 3c we simulate the potential profile of a double junction channel, consisting of regions doped at $10^{19}/cm^3$ at the left/right sides, but left undoped in the middle. With the dashed lines we show where the intrinsic region resides, having widths of $L_i$=10 nm (red lines), 20 nm (blue lines), and 30 nm (black lines). The solid lines show the self-consistently extracted potential profiles for each case, assuming p-type bulk Si density-of-states. Figure 3d shows the same profiles, but in this case the doping in the left/right regions is raised to $5 \times 10^{20}/cm^3$ (the corresponding elevated Fermi level is depicted in Fig. 3d). Clearly, appropriate barriers for energy filtering are formed, with their shape and height controlled by the dopant values and intrinsic region length. In the rest of the paper, for simplicity, we assume that the potential profile in the undoped regions begins from the edge of the doped region, and ends to the edge of the barrier region. (In principle, however, the potential barrier can be formed in a n++/i/n++ structure without the need of different barrier material itself). Due to the large differences in the doping, the depletion region indeed is shifted mostly in the undoped region, so it does effectively begin from the edge of the doped region. Note that the word 'depletion' throughout this work denotes both dopant and majority carrier depletion (compared to the n++ region) as in our previous works in Ref. [48, 49]. In principle, the middle-region barrier can be optimized by varying the dopant distribution accordingly, but in this work we assume intrinsic regions and provide the foundations of the design principle. We, thus, refer to these regions interchangeably as 'dopant-depleted', 'clean', or 'intrinsic'.

The TE coefficients in the case of the three-region structure can be computed by combining the individual coefficients of the three regions (well-W, intrinsic-i, barrier-B) as:

$$\frac{v_{tot}}{\sigma_{tot}} = \frac{v_W}{\sigma_W} + \frac{v_i}{\sigma_i} + \frac{v_B}{\sigma_B}, \tag{8a}$$

$$S_{tot} = \frac{S_W v_W + S_i v_i + S_B v_B}{v_W + v_i + v_B}. \tag{8b}$$

In a similar manner to the wells and barrier regions, the conductivity and Seebeck coefficient of the dopant-depleted regions (labeled 'i' for intrinsic from here on), can be extracted by:

$$\sigma_i = \int_{V_i}^{\infty} \tilde{\sigma}_i(E) dE, \tag{9a}$$

$$S_i = \frac{1}{\sigma_i} \int_{V_i}^{\infty} \tilde{S}_i(E) dE. \tag{9b}$$

In general, these quantities have a spatial dependence as the band edge changes in the 'i' region, but in most of the analysis below we consider an average band contribution, with that band edge located at mid-energy $V_B/2$ unless specified otherwise. (We have investigated various cases in our model, i.e. several band edge positions, each providing slightly different outcomes that do not change the foundation and advantages of our design). We also note that the models described by Eqs. 1, 3 and 8 are strictly valid for 1D periodic systems. Nanocomposites, on the other hand, are described by a 3D aperiodic geometry, and the complexity of the transport paths is such that would not allow us to map the 3D onto 1D paths beyond a first order estimation. The design we propose requires the current to flow normal to the wells and the barriers, rather than in parallel to them – i.e. the 2D geometry in Fig. 3a consists ideally of columnar grains extending into the page, as in Ref. [48]. Nanocomposites are also subject to geometry variations, and superlattice 1D geometries can also be considered as a limiting case for a nanocomposite system. Thus, we argue that Eqs. 1, 3 and 8 (with the volume fraction included rather than the length of the regions), are at first order applicable to nanocomposite/nanocrystalline materials as well.

Thermoelectric coefficients in the structure with dopant-depleted regions: The TE coefficients for the '3-region' structure are shown by the blue lines in the third column of Fig. 4, Fig. 4g-4i. Again, for comparison we also show the pristine material properties by the black-dashed line, and re-plot the result for non-relaxing wells of the second column by the red-dashed lines. The middle, dopant depleted intrinsic region is assumed to be of width $W_i$ = 5 nm here, and the TE coefficients are computed using Eqs. 8-9 above. The middle region reduces the Seebeck coefficient compared to the non-relaxing wells of the second column (blue versus red-dashed lines in Fig. 4h), but the dopant-free region strongly increases the conductivity of the overall domain (Fig. 4g). The carriers can now flow more easily under the weaker phonon-limited scattering conditions that prevail in the dopant-depleted regions. These regions geometrically occupy a significant volume of the structure even when having a narrow $W_i$ = 5nm width. Thus, the overall conductivity acquires a significant phonon-limited part with higher mean-free-paths, rather than an ionized impurity scattering limited part with much lower mean-free-paths. A significant power factor improvement is then achieved in this case as shown by the blue line in Fig. 4i.

Designing optimal attributes for the 3-region structure – allow thermal conductivity variations: Finally, we add another component to the design of the material, which brings an independent improvement in the Seebeck coefficient without first order changes in the conductivity. When different thermal conductivities ($\kappa$) exist in the different regions, the overall Seebeck coefficient can be generalized to (see derivation in the Appendix) [27, 37, 64]:

$$S_{tot} = \frac{S_W v_W / \kappa_W + S_i v_i / \kappa_i + S_B v_B / \kappa_B}{v_W / \kappa_W + v_i / \kappa_i + v_B / \kappa_B} \tag{10}$$

Here, we assume a ratio of $\kappa_W/\kappa_B = 5$ (and assume $\kappa_i = \kappa_W$), which is a reasonable ratio between the conductivities of grains and grain boundaries, for example. An additional increase in the Seebeck coefficient is achieved as shown in Fig. 4k (green lines vs blue-dashed lines). This new component leads to a larger power factor as shown in Fig. 4l. Here we included the TE coefficients from the second and third columns of Fig. 4 in dashed-lines for comparison. Overall, by considering these design strategies, a power factor increase of ~5× compared to the bulk material can be achieved. Note that such: i) dopant-depleted intrinsic regions, ii) semi-relaxation of energy, iii) undoped barriers, and iv) thermal conductivity variations, were used to explain experiments that measured very high power factors even up to ~22 W/mK$^2$ [48, 49, 51]. Below, we push the limits of what power factors can be achieved with this design principle.

Increasing the design parameters $W_i$, $V_B$, and $\kappa_W/\kappa_B$: To demonstrate the full potential of this design, we now proceed by increasing at a higher degree (but still under practically achievable limits) the parameters that allowed for the power factor improvements. We examine increases in $W_i$, $V_B$, and $\kappa_W/\kappa_B$ one by one, and finally combine all three of them together. The calculated results are shown in Fig. 5, where row-wise, as earlier, we show the TE coefficients $\sigma$, $S$, and $\sigma S^2$, and column-wise the influence of increasing $W_i$, $V_B$, and $\kappa_W/\kappa_B$, and all three simultaneously, respectively. In the sub-figures of the first three columns we still show for comparison the lines from the previous investigation in Fig. 4l, which indicate the effect on the TE coefficients for each design ingredient (dashed lines). In the first column, by increasing $W_i$ from 5nm to $W_i = 10$ nm (solid-blue lines), i.e. the length of the intrinsic regions is now longer, the conductivity increases significantly (Fig. 5a). Now a larger area of the material is composed of dopant clean regions, in which transport is phonon-limited with longer MFPs. The Seebeck coefficient is essentially not changed (Fig. 5b), despite the increase in the conductivity, because the elongated intrinsic regions raise the overall well band edge $E_C$ compared to the shorter $W_i$ case (see Fig. 3b-3d). This means that the $\eta_F = E_C - E_F$ is larger overall (in absolute terms), which tends to increase the Seebeck coefficient, thus mitigating the natural drop in the Seebeck when the conductivity is increased. Overall, therefore, the power factor largely increases to values ~13 mW/mK$^2$, dominated by the increase in the conductivity. Importantly, the power factor maximum is achieved for slightly smaller densities, following the shift of the conductivity to lower densities, which is easier to achieve experimentally.

Next, we examine the increase in the barrier height $V_B$ from $V_B = 0.15$ eV to $V_B = 0.25$ eV. The results are shown in the second column of Fig. 5, (Fig. 5d-5f) by the red-solid lines. The conductivity is significantly reduced up to the high densities of $10^{20}$/cm$^3$, where the Fermi level $E_F$ has still not yet reached the $V_B$. At higher densities, when the $E_F$ overpasses $V_B$, the conductivity is recovered to the pristine values, even slightly higher (with this slight increase depending on the details of the dopant-depleted regions $W_i$ and how their consideration is accounted for). The Seebeck coefficient, as expected, is largely increased (Fig. 5e), which makes the power factor to also largely increase (Fig. 5f). In this case, however, the power factor peak is shifted towards the higher densities, again following the shift in the conductivity.

The next step is to examine the increase in the ratio $\kappa_W/\kappa_B$ from 5 to 10 in the third column of Fig. 5, given by the green solid lines in Fig. 5g-5i. Independently of the conductivity, the Seebeck coefficient is improved, which reflects on PF improvements in Fig. 5i.

Finally, in the fourth column of Fig. 5, magenta lines in sub-Fig. 5j-*l*, we consider these larger increases of all the parameters together, i.e. increase the parameter values to $W_i = 10$ nm, $V_B = 0.25$ eV, and $\kappa_W/\kappa_B = 10$. Here we also kept the solid lines from each previous individual cases for reference. In Fig. 5j, the conductivity is benefitted significantly from the presence of the dopant free regions, and it is shifted to an intermediate region between the $W_i = 10$ nm alone (blue line), and the $V_B = 0.25$ eV alone (red line). Carriers still need high $E_F$ levels to be able to overpass the increased $V_B$, which allows for large Seebeck improvements as seen in Fig. 5k. Putting it all together, the increase in the power factor is quite substantial, reaching incredibly high values of >20 mW/mK$^2$ (magenta line in Fig. 5*l*), a factor of ~15× over the bulk value where we started from (dashed black line).

These are some very high values predicted by our model and simulations. Our model is simple in considering transport, and has also considered some idealized assumptions, such as energy non-relaxing transport and control on doping regions. However, we need to stress that even if those idealized conditions are not met in reality, there is still a lot of room for the power factor to be improved substantially over the bulk values. In any case, an efficient TE material can still be realized with even 5× improvement in the power factor. It is also quite remarkable that we consider such high barriers $V_B = 0.25$ eV, and the power factor still peaks at densities around $10^{20}/cm^3$, which are still achievable with current technologies for most TE materials. Thus, considering: i) this 'clean-filtering' approach, where the barrier is dopant-free, as well as ii) the support in the conductivity of the well from the dopant-depleted regions, a high conductivity is achieved at high Seebeck regions and incredibly high PFs can be realized.

Increasing $W_i$, $V_B$, and $\kappa_W/\kappa_B$ to the extreme for reaching the power factor upper limit: Next, we go one step further and consider extreme conditions under which the three parameters $W_i$, $V_B$ and $\kappa_W/\kappa_B$ are increased to the very high values of $W_i = 15$ nm, $V_B = 0.3$ eV, and $\kappa_W/\kappa_B = 15$. In the extreme $W_i = 15$ nm case, the entire well is dopant depleted (as $L_W = 30$ nm), and it is assumed that all dopants reside in a delta-function positioned in the middle of the well. In practice, a small heavily doped region can exist, which is there to supply the mobile carriers, without interfering with transport, an illustration of the modulation doping technique [37, 40]. The TE coefficients of some combinations of these values are shown in Fig. 6a-6c, indeed indicating incredibly high values of beyond 30 mW/mK$^2$. (The magenta-dashed line is the same as the magenta-solid line from Fig. 5 for comparison). It is quite instructive to show the mobility of these structures in Fig. 6d. The red-dashed line shows for reference the bulk phonon-limited mobility and the black-dashed line the bulk phonon plus ionized impurity scattering mobility, as in Fig. 1a. The proposed designs have much lower mobility for lower densities compared to both the phonon-limited and the phonon + IIS limited bulk mobilities. At higher densities, however, they all approach very closely the bulk phonon-limited mobility, and significantly overpass the bulk IIS-limited mobility. Thus, the materials, at high densities, owing to the dominant presence of the large dopant-free regions, are overall phonon scattering limited.

<u>Thermionic emission benefits:</u> In the second part of Fig. 6, (Fig. 6e-h), we repeat the same calculations as in Fig. 6a-d, but in this case we consider thermionic emission over the barrier, as shown in the left panel of Fig. 2e, and described by Eq. 7. The simple assumption of thermionic emission through a thin barrier allows the charge with energies higher than $V_B$ to flow 'freely' over the barriers and it is quite advantageous at higher $V_B$, which otherwise introduce strong resistance to the current. Thus, in this case the conductivities are much higher compared to the previous scenario (Fig. 6a vs Fig. 6e), the mobilities are higher (Fig. 6d vs. Fig. 6h), which doubles the power factor as well (Fig. 6c vs Fig. 6g) to extremely high values of beyond 60 mW/mK$^2$. Of course this is an overestimated value, which will drop once we consider interface resistance or the resistance that arises through quantum mechanical well/barrier momentum state mismatch. This mismatch will be is stronger with the barrier height as well. However, we examine the validity of this further below and point out that thin barriers where thermionic emission prevails can in general provide higher power factors.

We now devote the next part of the work to examining how realistic the conditions that we impose for obtaining such power factors are, using more advanced simulations. Specifically, we examine the position of the Fermi level at such high doping densities, the validity of the non- or semi-relaxation of current energy in our designs, the validity of thermionic emission together with the potential role of well/barrier state mismatch on the interface resistance, and the validity of the higher conductivity in SLs of n++/i junctions compared to uniformly doped channels. We also discuss the ideal volume fraction of the filtering barriers and potential practical implementations.

## IV. Discussion

<u>Doping level and the Fermi level position:</u> A major design ingredient in the design is the existence of a dopant-depletion region, whereas all the required doping resides into a small region in the middle of the grain. Thus, it is important to have an estimate where the $E_F$ resides with respect to the band edge $E_C$ in the doped region. We already expect to operate in the highly degenerate conditions (locally in the middle core of the wells, with the $E_F$ residing high into the bands), although in the barrier regions of course the $E_F$ will be below the band profile. In the dopant-depleted regions, the band will cross the $E_F$. The question we essentially want to answer here is: how high can we dope the middle region, or how high will the $E_F$ be if a certain mobile carrier density needs to be achieved in the entire material region, given that it will only be supplied from the central/core region (blue-colored regions in Fig. 3a)? Ultimately, as shown in Fig. 3c and 3d, the built-in barriers are formed in the undoped region, which is largely extended since the depletion region is preferentially placed in the intrinsic region of the n++/i junction. On the other hand, the depletion region in the highly doped core is much narrower due to the very high doping (even just a few nanometers – see Fig. 3c-3d for the non-flat band regions left/right of the dashed lines). Essentially, those narrow regions are depleted of the same amount of carriers that appear in the entire undoped barrier region, $W_i+L_B$ combined. The illustrations of Fig. 3c and 3d are extracted from self-consistent Poisson equations, and clearly indicate that just a few nanometers of highly doped regions are enough to support the depletion and barrier regions with the necessary carrier density needed for the power factor to peak to very large values. The higher the doping of the highly doped regions, the smaller the extension of the depletion into the doped regions and the larger the depletion regions in the intrinsic/barrier regions (Fig. 3d).

In Fig. 7a we show the PF computations for the case of the different depletion region sizes considered, i.e. $W_i$ = 5 nm (blue line same as in Fig. 4i), 10 nm (red line), and 15 nm (entirely dopant depleted wells, with delta-function shaped core distributed doping, green line). Interestingly, the larger the dopant-depleted region, the higher the power factor, but the smaller the required doping density in the middle of the well to achieve maximum PF, which would be easier to achieve experimentally anyway. Figure 7b shows the position of the Fermi level in the middle region for the different cases of dopant-free region widths, in the structure where the barrier height is $V_B$ = 0.15 eV as depicted. With the black-dashed line we show the $E_F$ position with respect to the band edge $E_C$ = 0 eV for the pristine uniformly doped channel case. Note that we consider p-type Si DOS parameters as high PFs are achieved for p-type Si [48], although we invert the bands to talk about conduction bands, as it is easier to perceive a flow over barriers than below barriers. As the regions where dopants are allowed (blue-colored regions in the middle of the wells in Fig. 3a) are reduced ($W_i$ increases), a higher $E_F$ positioning (higher center region doping) is needed to achieve the same carrier density. Around the density needed for maximum power factors $5\times10^{19}/cm^3$, the $E_F$ is pushed into the band by 0.1 eV, almost $3k_BT$ higher than what would supply the same density if the structure was uniformly doped. The last case, $W_i = 15$ nm, leaves no room for doping to be placed into, as the length of the well region we consider is $L_W = 30$ nm. However, as it was stated above, it should be understood, that there could be a narrow region of few nanometers of highly doped regions in excess, i.e. the well size would have been slightly larger. These are a small part of the well volume, still, however, enough to provide the necessary density in the intrinsic regions and the barrier regions.

The high doping in the central region then serves an important role, as it pushes the bands low for high velocities to participate in transport. The fact, however, that a narrow volume of the material is responsible for suppling the entire doping required (from $10^{19}cm^3$-$10^{20}/cm^3$), requires that the $E_F$ levels increase substantially, to ~0.1 eV into the bands, with the doped regions reaching doping values at the levels of $10^{20}cm^3$ - $5\times10^{20}/cm^3$ (to see this, one can find the $E_F$ in Fig. 7b corresponding to the maximum PF of Fig. 7a, and then project horizontally that level until it meets the dashed line). Interestingly, from Fig. 7b we can extract that to achieve $5\times10^{19}/cm^3$ overall concentrations in the $W_i = 15$ nm structure, concentrations of only ~$2\times10^{20}/cm^3$ in the middle blue-colored regions are needed. As indicated by the narrow depletion lengths of the heavily doped regions in Fig. 3c-3d, a few nanometers of such central region would suffice. Such dopant values are realistic within current technologies, for example in Si, and even more importantly for nanocrystalline Si, where dopant solubility thresholds allow for concentrations as large as $5\times10^{20}$ and $2\times10^{21}$ $cm^{-3}$ using boron and phosphorus, respectively [65]. The simulations here are performed using Si parameters, but the concept we present is applicable in general to other materials as well.

<u>Controlling the energy relaxation:</u> An important aspect of the design is the reduced energy-relaxation in the well region, such that the high Seebeck coefficient of the barriers is transferred throughout the material. Thus, it is useful to point out what determines energy relaxation, what is the degree of energy relaxation under realistic conditions, as well as to propose some practical design directions which provide control over relaxation. Regarding geometrical features for the size of the well, several works by us and others point out that energy relaxation is prevented, or mitigated, when the well sizes are in the range of the energy relaxation length, $L_W \sim \lambda_E$ [26, 27, 64]. Energy relaxation in semiconductors is dominated by inelastic scattering processes, primarily electron-optical phonon scattering. In Si, for example, the electron-optical phonon mean-free-path is around $\lambda_E \sim 15$ nm, which results in wells sizes of $L_W \sim 50$ nm to exhibit semi-relaxation of the current energy [27, 64]. Well sizes of the order of $L_W = 30$ nm will only exhibit some degree of relaxation as we show below. However, other than the well size, there are other parameters that can contribute to reduced carrier energy relaxation in the design proposed, and they are discussed below.

Figure 8a shows non-equilibrium Green's function (NEGF) quantum transport simulations in SL channel geometries with semi-relaxing $L_W = 50$ nm wells (Fig. 8a). For this, we have used our in-house 2D simulator that we have developed with details explained in Refs. [26, 64, 66]. We include the effect of electron-phonon scattering with different common phonon energies from $\hbar\omega = 0.02$ eV to 0.09 eV (leading to different $\lambda_E$), for an example case where $V_B = E_F + 0.05$ eV. The colormap indicates the current flow in space and energy, with yellow indicating high current, whereas the curved lines indicate the energy of the current flow <E>, which determines the Seebeck coefficient as in Eq. 5. From our NEGF simulations, this is extracted by scaling the current $I_{ch}(E,x)$ at all spatial points as:

$$\langle E(x) \rangle = \frac{\int_0^\infty I_{ch}(E,x) E dE}{\int_0^\infty I_{ch}(E,x) dE}. \tag{11}$$

Carriers flow over the barrier and then relax into the well in a distance determined by the relaxation length $\lambda_E$. In NEGF we can control $\lambda_E$ by adjusting the electron-phonon scattering strength, $D_O$, and the phonon energy, $\hbar\omega$. The short horizontal lines in the second well show the level where the <E> of the current flow will reside at, in the case where the barriers are absent, i.e. the level at which the current energy relaxes in the pristine material. For all values of $\hbar\omega = 0.02$ eV to 0.09 eV examined (energies for typical TE materials), the current is not relaxed at the pristine well level, but is higher, on average in between the energy levels observed on the top of the barriers and those in the pristine wells.

Indeed, in Fig. 8b we show the Seebeck coefficients of these channel materials with varying $\hbar\omega$ (red line) and varying phonon strength $D_O$ (blue line), but plotted as a function of the $\lambda_E$ that they correspond to (again typical values for TE materials and semiconductors). The $\lambda_E$ is extracted by fitting the relaxing $<E>$ by an exponential function $e^{-x/\lambda_E}$ for a structure with a single barrier as indicated in the inset of Fig. 8b. The dashed-flat lines show the Seebeck coefficients of the pristine channel $S_W$, and a channel consisting of one large barrier extending across its entire length, $S_B$, for the $\hbar\omega = 0.06$ eV (as in Si) – for clarity we do not show the lines for the uniform channels for the rest of the phonon energies, but we note that our calculations indicate a ~20% variation at most for the other phonon energies. The well size $L_W \sim 50$ nm of this example dictates semi-relaxation for the Si case, and thus the overall SL Seebeck coefficient resides in the upper half of the allowed range between the pristine values $S_B$ and $S_W$ (despite the fact that the barriers occupy a much smaller portion of the channel compared to the wells). Materials with weak electron-optical phonon scattering, or materials with large optical phonon energies (right side of Fig. 8b), on the other hand, have weaker relaxation, with the Seebeck coefficient being even closer to that of the pristine barrier material $S_B$.

Evidently, whether $\hbar\omega$ or $D_O$ is responsible in altering $\lambda_E$, the actual relaxation and overall Seebeck coefficients are different, even at the same $\lambda_E$ (simply, the red/blue lines in Fig. 8b differ). The overall $\lambda_E$ and Seebeck coefficient is mostly linear with electron-phonon coupling strength $D_O$ (blue line). On the other hand, for very small, or very large $\hbar\omega$ values, even at the same expected $\lambda_E$ as extracted from the single barrier cases, once a SL channel is formed, the actual relaxation is less, and the Seebeck coefficient is higher (higher red versus blue line). The reasons are the following: *i) Small phonon energies:* When the phonon energy $\hbar\omega$ is small, the $\lambda_E$ decreases, but scattering approaches the elastic limit, and thus relaxation saturates, and is even suppressed for $\hbar\omega \to 0$. *ii) Large phonon energies:* For an electron at energy $E$ to emit an optical phonon and relax at a lower energy, an available empty state has to exist at energy $E - \hbar\omega$. As electrons primarily flow over the barriers and into the wells, they would tend to relax around the Fermi level $E_F$ (which in this case is placed into the bands at degenerate conditions). If, however, $V_B$ and $E_F$ are placed close in energy such that $\hbar\omega > V_B - \max(E_F, E_C(x))$, then energy relaxation is suppressed due to the absence of empty states below the Fermi level, also evident by the black-dashed line in Fig. 8a for the largest phonon energy we considered. Thus, the simplified picture of $\lambda_E$ extraction from a single barrier is no longer valid in SL structures, which tend to exhibit reduced energy relaxation, an advantageous observation. The point here, however, is that starting from the optimal conditions of a semi-relaxed channel ($L_W$ = 50 nm, $\hbar\omega$ = 0.06 eV, $\lambda_E \sim$ 15 nm), and by varying the phonon strength and the phonon energies such that $\lambda_E$ changes by 30% in either direction, still the Seebeck coefficient tends to move mostly towards that of the barrier, rather than that of the well in most of the parametric region.

Further on, to relate to the channels we have simulated earlier within the BTE and Si parameters, we have chosen a channel with $\lambda_E$ = 15 nm and $\hbar\omega$ = 0.06 eV (resembling Si), and performed simulations for different well sizes from the ultra-short $L_W$ = 2 nm, up to $L_W$ = 100 nm. Figure 8c shows the Seebeck coefficients for these structures (red line), whereas again the dashed-flat lines show the Seebeck coefficients of the pristine channel $S_W$, and the channel consisting of one large barrier $S_B$. (The semi-relaxing $L_W$ = 50 nm channel we considered earlier has a Seebeck coefficient which is in the middle of the two limiting cases). Clearly, however, for the $L_W$ = 30 nm channels that we considered in the BTE simulations earlier, the Seebeck coefficient is $S_{SL} \sim$ 180 μV/K, which is much closer to the $S_B \sim$ 210 μV/K, rather than the $S_W \sim$ 85 μV/K. This justifies our considerations of non-relaxing current energy in the wells. (Note that in practice the optimal power factor conditions are found when some relaxation is present, such that the barriers are spaced as far as possible to reduce the density of interface resistances, but as close as possible to prevent relaxation – we discuss interface resistance reduction directions further below). Interestingly, the magenta-dashed line shows an analytical calculation of the SL Seebeck coefficient if the well/barrier are considered independently (as in Fig. 2b). In this case much lower Seebeck coefficients are achieved for the SL.

Reduced relaxation and interface resistance in the 3-region structures: It is also worth discussing a few other things that point towards reduced energy relaxation in the wells of the proposed design. For example, the shape of the band edge $E_C$ in the dopant-free regions, as shown in Fig. 3c and 3d, with oblique band edges: i) provides higher Seebeck coefficient due to the higher average $E_C$, and ii) further reduces the availability of empty states at lower energies for electrons to relax their energy into. This reduces the energy window for optical phonon emission to happen and thus, reduces the energy relaxation. This is clearly indicated in Fig. 9a, where we have performed NEGF simulations using electron-optical phonon scattering alone and altered the barrier potential to the oblique shape as a first order approximation of what shown in Fig. 3b. As the sidewalls of the barrier become more and more oblique, the energy in the wells begins to raise. The oblique sidewalls, however, tend to reduce the energy of the current flow above the barriers, but finally modest improvements to the overall Seebeck coefficient are observed (Fig. 9c). Furthermore, in general, highly doped regions (in the well core), also push the energy of the current upwards as lower energies scatter more effectively off dopants (evident from the anisotropic Brooks-Herring scattering model [56]). Thus, in the structures proposed, the combination of: i) the chosen length for the well to be in the order of $\lambda_E$, ii) the 'oblique' band shape in the dopant-free regions, and iii) the doped core, allows for a significant degree of energy non-relaxing transport in the wells. The important point, here, is that the reduced energy relaxation in the wells is justified also by consideration of quantum transport simulations. This justifies the choice of the beneficial non-relaxing consideration in the BTE simulations earlier.

As a side note, we have shown in the past that nanoinclusions (NI) in the well regions, having barrier heights up to the $V_B$ of the superlattice, can push the energy of the current flow upwards and further increase the Seebeck coefficient, also allowing for small, but noticeable power factor improvements [26, 66]. That would be something to also provide lower thermal conductivities, with additional benefits for *ZT*.

Oblique sidewalls reduce interface resistance: It is also quite interesting to observe the conductance of the channels in Fig. 9a, plotted versus the sidewall distance *d*. The reduction in the 'local' Seebeck in the barriers in Fig. 9c at first when using oblique profiles is a signal of reduction in well/barrier interface resistance, a consequence of better well/barrier state matching. Indeed, our simulations show that the quantum mechanical transmission over the barriers increases in the oblique cases compared to the sharp barriers, and quantum reflections/oscillations are smoothened out. Due to this, the conductance is improved at first instance by ~20% (Fig. 9b). For larger $d$, the conductance remains almost constant, whereas the Seebeck coefficient increases from contributions in the wells. Overall, the introduction of the sidewalls increases the power factor monotonically, up to values of ~30% (Fig. 9d).

The validity of thermionic emission (Fig. 10): When it comes to the behavior of carriers over the barrier, we have shown up to now that 'dopant-clean' barriers, with oblique sidewalls for reduced interface resistance, and/or thermionic emission from the wells over the barriers are important ingredients for the design. Here, we examine the validity of the thermionic emission, again using NEGF simulations, this time including electron-acoustic phonon scattering only (elastic scattering to isolate the effect of carrier relaxation on the barrier from inelastic relaxation processes into the wells). We simulate a channel with a single potential barrier in the middle, and vary the length of the barrier $L_B$ from $L_B = 100$ nm (taking over the entire channel) to $L_B = 5$ nm and then to zero, i.e. the pristine channel case (as illustrated in the inset of Fig. 10). In Fig. 10 we plot the energy resolved transmission function, defined as $Tr = (h/q_0^2)I_{ch}/(f_1-f_2)$, where $I_{ch}$ is the NEGF extracted current, $h$ is Plank's constant, and $f_1$ and $f_2$ are the Fermi-Dirac distributions of the left/right contacts, respectively [26]. All these quantities are energy dependent. The $Tr$ is directly related to the transport distribution function in the BTE, and has a linear dependence in energy in the case of acoustic phonon scattering for a single subband [66, 67]. That linear dependence is captured in the (multi-band) NEGF simulations for the pristine channel (brown line), as well as the long barrier channel (purple-dashed line), with the initial point being the band edge, i.e. 0 eV in the pristine case, and $V_B = 0.05$ eV in the long barrier case.

As the barrier length $L_B$ is scaled, however, there is a clear shift at energies after $V_B$ towards the $Tr$ of the pristine channel. The 'jump' in the $Tr$ after $V_B$ in the shorter $L_B$ = 5 nm barrier channel, clearly indicates that carriers 'see' the barrier, but for energies above the barrier they have a $Tr$ more similar to that of the well. This would be an indication of ballistic thermionic emission, in which case the carriers do not relax (at a large degree) on the bands of the barrier, i.e. overall they do not acquire the low velocities at the top of the barrier, but propagate with the well attributes. The gradual change of the black line $Tr$ in Fig. 10 towards that of the well (reaching at ~80% of that value within 0.01 eV after the barrier at $V_B = 0.05$ eV), could signal that some well/barrier state mismatch and quantum reflections are still present, adding to the interface resistance. We also note that the aforementioned energy window deviations from the pristine $Tr$ would increase as $V_B$ increases due to larger mismatch. However, any acquired slope in $Tr$ being larger compared to the one of the large non-thermionic well (brown line in Fig. 10), would be beneficial to the conductivity and the power factor. This effect would be reduced, however, when: i) the barrier sidewalls acquire a slope as explained above in the discussion of Fig. 9a, and ii) if the barrier material has much more transport modes compared to the well material, such that more momentum state matching is achieved [68]. This can be the case of a barrier material with much higher effective mass, for example. Thus, the optimal power factor of our design could be somewhere in between the ~30 mW/mK$^2$ and ~60 mW/mK$^2$ indicated in the two examples of Fig. 6.

High electronic conductivity in n++/i SL structures (Fig. 11): Since the major new element this work proposes is the introduction of the dopant-free region separating he highly doped core or the potential well and the barrier, Monte Carlo simulations were further performed to verify the outcomes of the simple model employed in the main part of this paper. We employ a home-developed 2D electron transport Monte Carlo code, and as a simple example we simulate three channels as indicated in Fig. 11a-c: i) a uniformly doped channel without barriers, ii) a uniformly doped channel, but with barriers, and iii) a channel where only the well regions are doped, and the barriers are not. The shape of the barriers is extracted from a self-consistent Poisson solver, corresponding to doped regions of $L_{\text{doped}} = 30$ nm at $N_D = 4\times10^{19}$/cm$^3$, and $L_{\text{undoped}} = 20$ nm. P-type Si density-of-states were used. It is not our intention here to perform a complete study using the numerically expensive Monte Carlo simulation, but to indicate that indeed doping variation can mitigate the conductivity reduction due to the introduction of potential barriers. Thus, considering the simplicity of non-relaxing transport, we only consider elastic scattering (acoustic deformation potential scattering and IIS). We report on the electronic conductances of these three structures: i) The uniformly doped channel, has $G = $ 9.36×10$^{-6}$ S. ii) The conductance of the uniformly doped channel, but with barriers as well, suffers a large reduction down to $G$ = 1.78×10$^{-6}$ S, a factor of 5.3×. iii) 'Cleaning-up' the dopants from the barrier regions (as is the actual case for the n++/i self-consistent simulation here), helps the conductance to increase to $G$ = 5.92×10$^{-6}$ S, recovering to ~60% of that of the uniform channel. The behavior is clearly observed in the energy resolved current of electrons in the Monte Carlo simulator, shown in Fig. 11d for the three cases. The current in the uniformly doped SL (red line) case picks up after the barrier, but even then remains lower compared to the pristine channel case (black line), as carrier on the band edge on the top of the barrier have lower velocities. On the other hand, the third case of the non-uniformly doped channel (blue line), picks also after the barrier, but it is significantly higher compared to the pristine case, contributing to increased conductivity. Finally, as the introduction of the barriers (by simple considerations) increases the average energy of the current <$E$> by 2.5×, a more than doubling of the power factor would be expected. The closest simulations that this can relate to in the earlier BTE simulations is what observed in Fig. 4i at the data points of the corresponding densities.

Intrinsic Seebeck of the boundaries and their volume fraction: The boundary regions (gray regions in Fig. 3a) are there to facilitate the creation of the dopant depleted regions (as for example in the nanocrystalline structures of Ref. [48, 49, 50, 51]). The boundary regions are also important in providing the barrier $V_B$, which will form when the boundary material and the potential well material have a band edge discontinuity, $\Delta E_C$). On the other hand, even if the boundary material does not have a significant $\Delta E_C$ (or even if it has a negative one), the presence of the junction between the heavily doped well core and the undoped/lightly-doped middle and boundary regions will electrostatically form an effective barrier $V_B$. Thus, the isolated 'intrinsic' Seebeck coefficient of the boundary material will only have a secondary effect. The primary determination of the Seebeck coefficient and PF will be determined by the effective built-in barrier (band edge discontinuity and electrostatics). Note that in the case where the junctions (blue/red regions in Fig. 3a) are achieved with selective local doping on monocrystalline materials, even in the absence of boundary regions large benefits will also be expected. Where boundaries, as a second phase, can provide an additional improvement to the overall Seebeck coefficient, is through variability in the thermal conductivity.

With regards to the volume fraction that the boundaries occupy, within the assumption that they remain undoped, our simulations show that moderate PF benefits are possible (even up to 2×) compared to the uniformly doped, pristine material, even if the boundaries occupy a larger volume fraction compared to the wells/grains. To obtain the very large PF improvements we present, however, the boundary regions need to be smaller than 5 nm in length, in order to mitigate electrical resistance and allow for a degree of thermionic emission (as in Fig. 10). On the other hand, boundary regions smaller than 3 nm will allow quantum mechanical tunnelling, and result in smaller temperature drops across them, both which reduce the Seebeck coefficient [24]. Thus, we suggest that the boundaries are optimally of the order of 5 nm thick. In the case of Si, for example, to achieve semi-relaxation of the current flow in the channel, the grain/well size should be in the order of $L_W$ = 30 nm - 50 nm. This leads to an optimal volume fraction ratio of at most ~25% for the boundary regions.

<u>Example of possible practical realization:</u> The simplest experiment to design and evaluate the potential of the 'clean-filtering' approach is to begin with two regions, and fabricate 2D superlattices formed of n++/i, or n++/n- junctions. In that case the barriers are formed in the intrinsic or lightly doped regions, which will be regions 'cleaner' of dopants, having phonon-limited mobility. Lithography can be used for the definition of windows through an oxide layer (grown by thermal oxidation, for example) on an SOI wafer, and shaped by lithography and etching to act as a mask for the doping process. Oxide windows, and hence the final doping concentration, can be arranged to form a 2D array using ion implantation (for example). As a next step, one can go even further by lithographically defining lines in the x- and y-directions to form a square 'net', and then dopant diffusion can create highly doped islands in the regions between the 'square net' lines as in the blue-colored islands of Fig. 3a.

# V. Conclusions

In conclusion, this work proposes a novel design direction which will allow nanostructured materials to deliver exceptionally high thermoelectric power factors, even more than an order of magnitude compared to the original material's corresponding values. The design is based on an extension and generalization of previously presented strategy that realized experimentally very large power factors (5× compared to optimized pristine material values) [48, 49]. In this work, it is shown that much higher power factors can be achieved once the grain/grain-boundary (well/barrier) design is properly optimized. Specifically, the proposed design utilizes energy filtering where carriers flow from heavily doped potential wells into undoped barriers (with a degree of thermionic emission) for reducing the barrier resistance that ionized impurities would have caused. Importantly, though, it introduces an undoped region, 'clean' of dopant impurities, that separates the core of the wells from the barriers. This essentially allows for higher carrier mean-free-paths and mobility in the wells (approaching phonon-limited), compensating by far the conductivity reduction caused by the barriers. The work also points out that other than the 3-regions (1-heavily doped well core, 2-intrinsic carrier path spacer, 3-potential barrier), an essential ingredient is that the energy of the carriers does not relax significantly in the well regions. It is shown, however, that this is the most probable and realistic case in the design we propose, as: i) the potential well length can be chosen such as to be in the same order as the energy relaxation mean-free-path, ii) the band edge shape of the intrinsic regions favors reduced relaxation, and iii) degenerate doping conditions also favor reduced energy relaxation. Thus, the design can provide exceptionally high power factors because it: i) reduces the resistance of the barriers, ii) transfers the high Seebeck coefficient of the barriers into the wells, and iii) allows very high conductivity in the wells. The latter is achieved by utilizing dopant-free intrinsic regions for transport, but with high energy carriers that provide high carrier mean-free-paths and phonon-limited, rather than ionized impurity scattering dominated mobilities. Although some of the parameters the simulations employ are relevant to Si, the design approach can be applied in general to other materials as well.

## Acknowledgements

This work has received funding from the European Research Council (ERC) under the European Union's Horizon 2020 Research and Innovation Programme (Grant Agreement No. 678763). We acknowledge Hans Kosina and Mischa Thesberg for helpful discussions.

# Appendix

Here we provide the theoretical proof of the equation for the Seebeck coefficient of a composite system, which leads to Eq. 3b in the main paper. The Seebeck coefficient of an arbitrary irregular system is defined as the weighted average of the local Seebeck coefficients along the length of the material ($L$), with the weighting factor being the lattice temperature gradient as:

$$S = \frac{\int_0^L S(x)(dT_L/dx)dx}{\Delta T} \qquad (A1)$$

where for the applied temperature difference $\Delta T$, it holds $\Delta T = \int_0^L (dT_L/dx)$. Here we assume an 1D channel material. We assume that the lattice temperature ($T_L$) varies according to a simple thermal circuit model. In this case, each of the two materials forming the composite system (barriers and wells) has a different temperature gradient across it depending on its thermal conductivity, as $dT_L/dx|_B$ for the barriers and $dT_L/dx|_W$ for the wells. The entire temperature drop across the material is then decomposed as:

$$\Delta T = \frac{dT_L}{dx}\bigg|_B L_B + \frac{dT_L}{dx}\bigg|_W L_W \qquad (A2)$$

where $L_B$ and $L_W$ are the total lengths of the wells and barrier regions. At an interface between different materials, the heat flux is conserved, thus using Fourier's law we have:

$$J_Q = -\kappa_B \frac{dT_L}{dx}\bigg|_B = -\kappa_W \frac{dT_L}{dx}\bigg|_W \qquad (A3)$$

Substituting (A3) into (A1), we reach:

$$S = \frac{S_B L_B \frac{J_Q}{\kappa_B} + S_W L_W \frac{J_Q}{\kappa_W}}{L_B \frac{J_Q}{\kappa_B} + L_W \frac{J_Q}{\kappa_W}} = \frac{\frac{S_B L_B}{\kappa_B} + \frac{S_W L_W}{\kappa_W}}{\frac{L_B}{\kappa_B} + \frac{L_W}{\kappa_W}} \qquad (A4)$$

In the case of a 3D material with irregularities in the distribution of the wells/barriers, for example as in a polycrystalline material of grains/grain boundaries, the lengths are replaced by the volume fractions of the different regions [37], i.e., the overall Seebeck coefficient of a 3D nanocomposite material is approximately the volume weighted average of the Seebeck coefficients of the constituent material phases as:

$$S = \frac{\dfrac{S_B V_B}{\kappa_B} + \dfrac{S_W V_W}{\kappa_W}}{\dfrac{V_B}{\kappa_B} + \dfrac{V_W}{\kappa_W}} \tag{A5}$$

Figure 1:

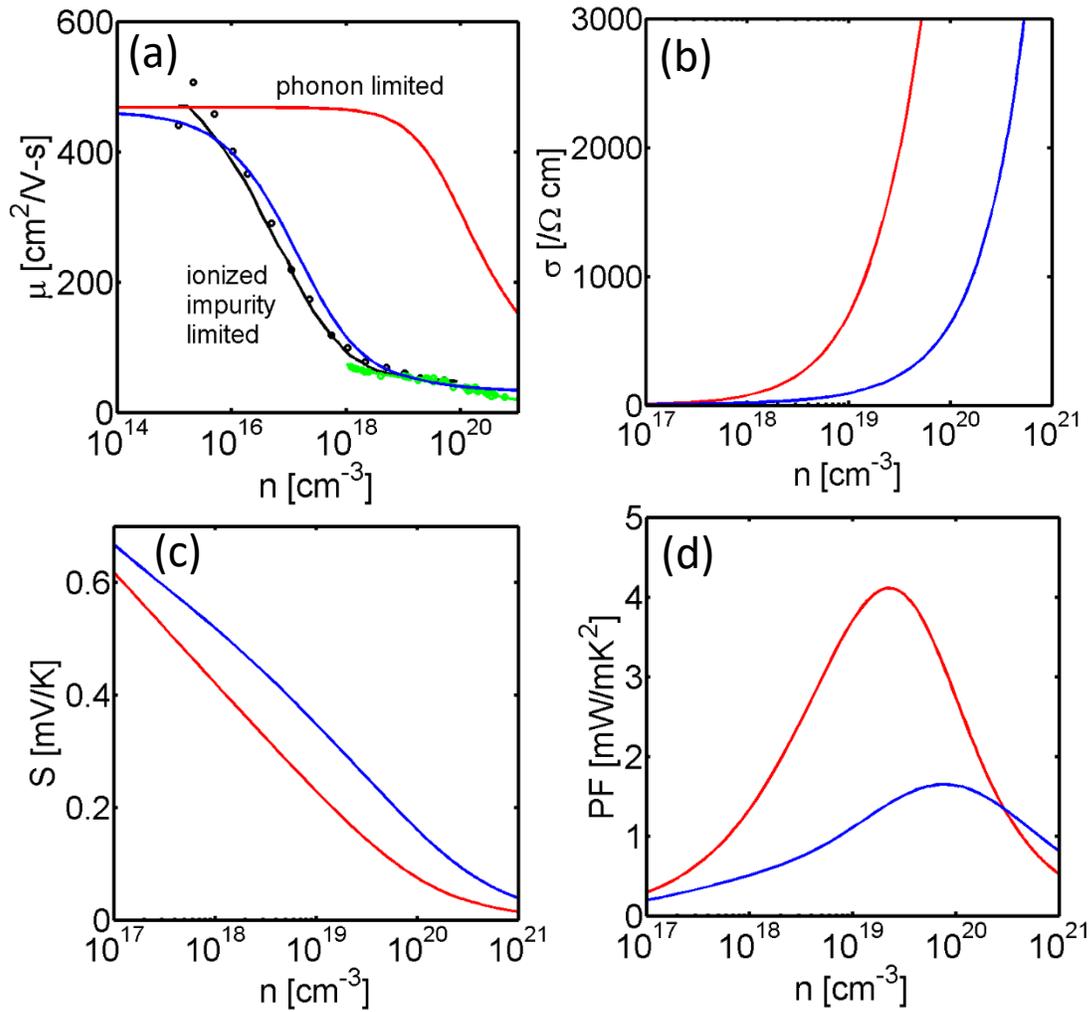

Figure 1 caption:

(a) The mobility of p-type Si from experimental data (by Jacoboni [59]– black dots and Masetti [61] – green dots), the calculated phonon limited mobility (red line), and the benchmarked phonon plus ionized impurity scattering from our semiclassical model (blue line). (b-d) The conductivity, Seebeck coefficient, and power factor for the cases of the phonon-limited (red lines) and phonons plus ionized impurity scattering (blue lines) considerations.

Figure 2:

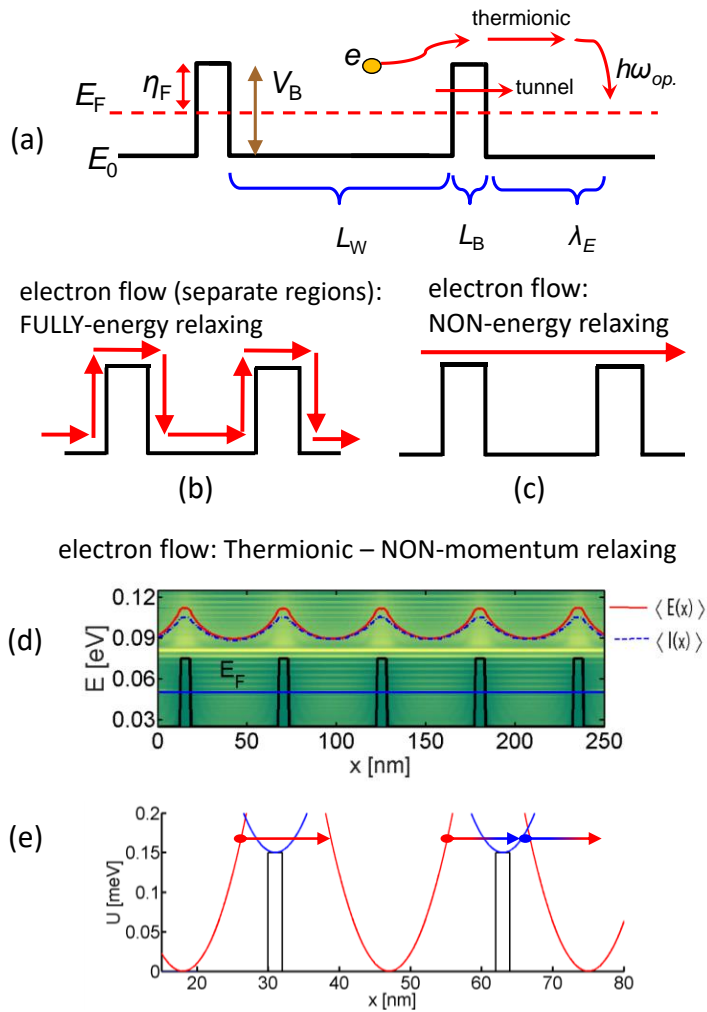

Figure 2 caption:

(a) Schematic of typical electron transport in a superlattice structure, where electrons flow over potential barriers and relax into potential wells. (b-c) Transport in a superlattice in the extreme cases where energy relaxation into the well after overpassing the barrier happens immediately and fully, and when energy relaxation does not happen at all. (d) Non-Equilibrium Green's function simulations for the energy of the current flow in a superlattice (red line), indicating that carriers (blue line) relax into the wells partially, depending on the sizes of the well regions. The yellow/green colormap indicates the current flow $I(E,x)$. (e) Schematics indicating situations resembling thermionic emission over the barrier (left), and emission over the barrier where carriers completely relax their momentum on the band edge at the top-of-the-barrier (right).

Figure 3:

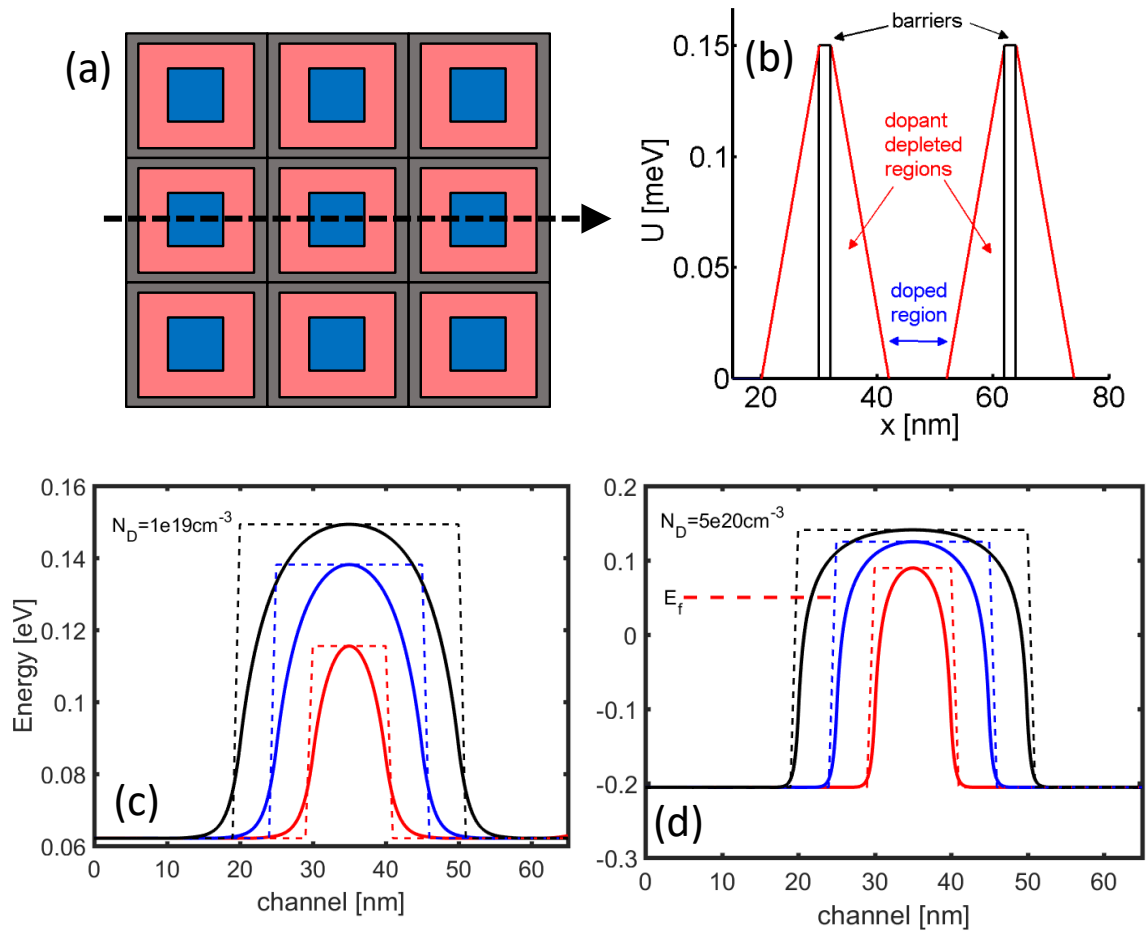

Figure 3 caption:

(a) The three-region structure design proposed in this work in a 2D top-down representation. The core of the well region (blue regions) is highly doped, the intermediate region between the core and the boundary (red region) is intrinsic (dopant-free), and the region in-between the wells is shown in grey. (b) A 'cut' through the dashed line of (a), indicating a simplification of the conduction band profile, with the doped regions, the intrinsic regions and the barrier regions indicated. (c-d) Band profiles after self-consistent solutions of the Poisson equation for doping values $N_D = 10^{19}$ /cm$^3$, and $5 \times 10^{20}$/cm$^3$. In each sub-figure, cases for intrinsic regions of length 10 nm (red lines), 20 nm (blue lines), and 30 nm (black lines) are indicated (doped/undoped region boundaries are shown by the dashed lines). The Fermi level in (d) is denoted by the red-dashed line.

Figure 4:

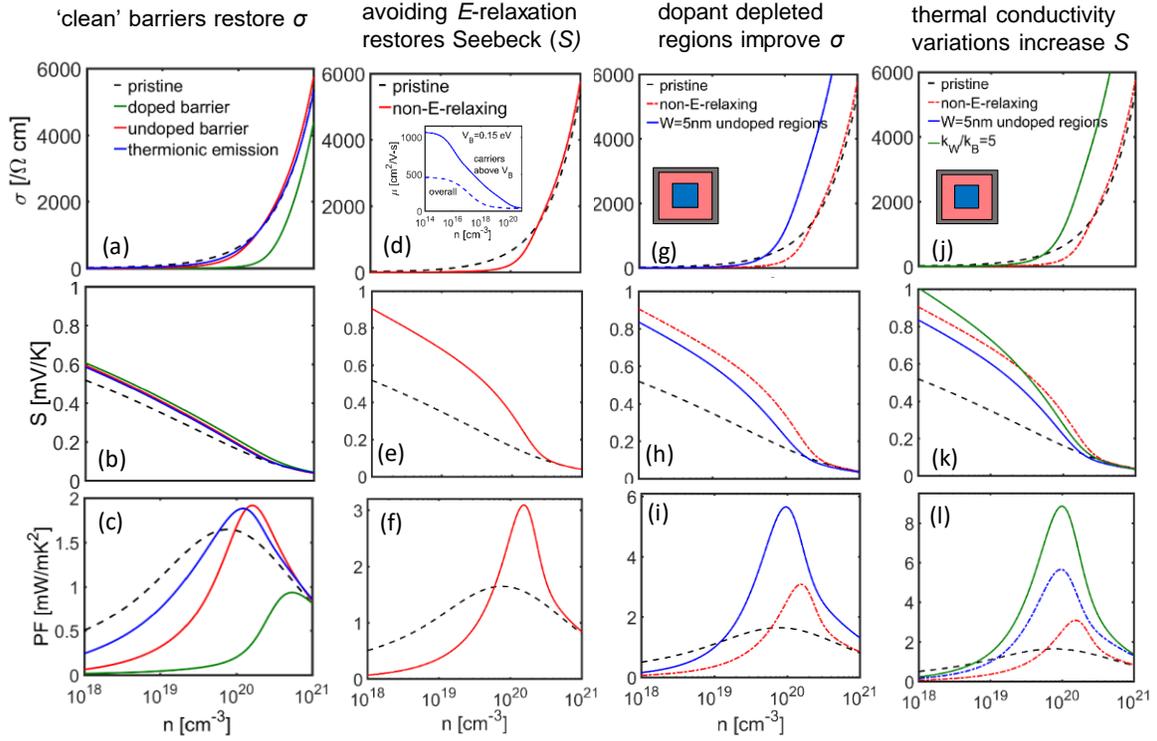

## Figure 4 caption:

The proposed design structure 'ingredients' one-by-one column-wise. Row-wise: the electrical conductivity, Seebeck coefficients, and power factors versus carrier density. In all cases, the black-dashed line shows the properties of the pristine flat band channel, without any barriers included for comparison. Column 1 (a-c): Comparison between the pristine case, and superlattices where the well/barrier regions are both doped and independent of each other (green line), and cases where the barrier is undoped (blue lines) and where thermionic emission prevails above the barrier (red lines). The barrier height is $V_B = 0.15$ eV. Column 2 (d-f): Comparison between the pristine case and superlattices when considering the undoped barrier case and completely unrelaxed current energies in the wells (red lines). Inset of d: The mobility of the isolated carriers that flow over the $V_B$ (solid line), compared to the mobility of all carriers (dashed line). Column 3 (g-i): Comparison between the pristine case and the case of undoped barriers, unrelaxed energies in the wells, and dopant-free regions of length $W_i = 5$ nm in the well next to the barrier regions (blue lines). In dashed-red lines the data from column 2 are repeated for direct comparison. Column 4 (j-l): Comparison between the pristine case and the case where the thermal conductivity is not uniform in the well and barrier regions such as $\kappa_W/\kappa_B = 5$ with all other parameters as in Column 3 (green lines). In dashed-red lines the data from Column 2 and in blue-dashed lines the data from Column 3 are repeated for direct comparison.

Figure 5:

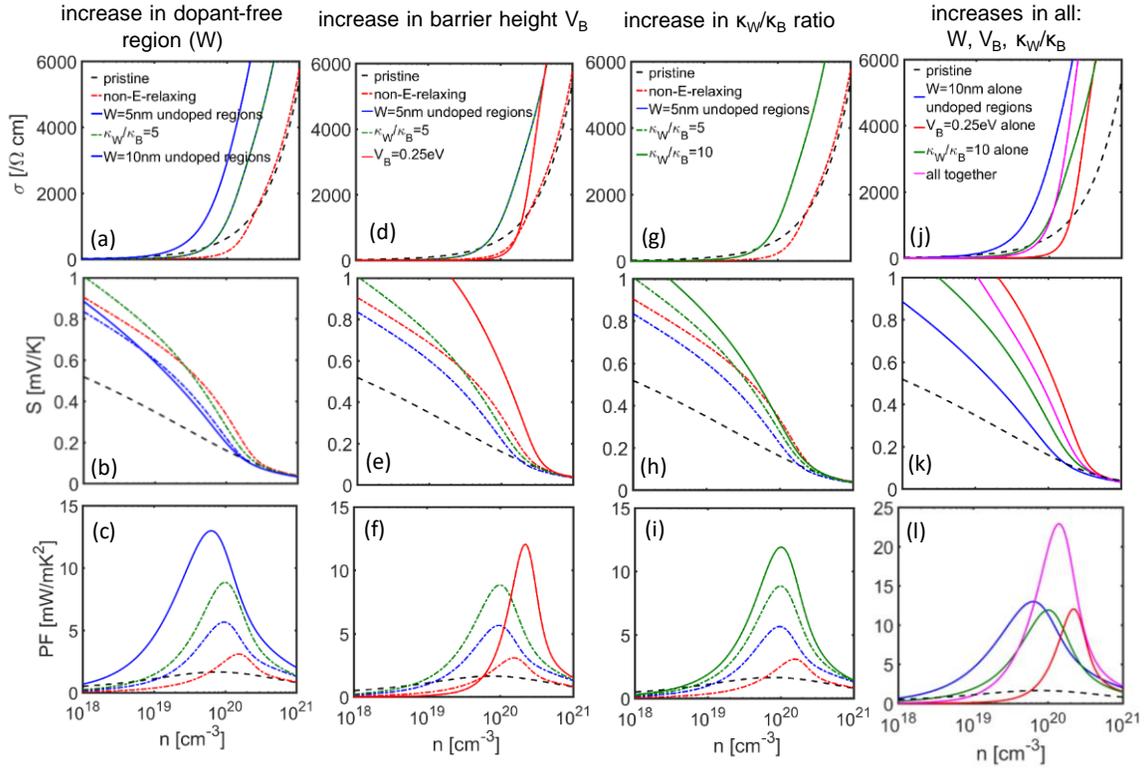

Figure 5 caption:

The proposed design structure 'ingredients' increased one-by-one column-wise to higher values. Row-wise: the electrical conductivity, Seebeck coefficients, and power factors versus carrier density. In all cases, the black-dashed line shows the properties of the pristine flat band channel, without any barriers included for comparison. The red-dashed, blue-dashed and green-dashed lines are the corresponding solid line data from Fig. 4 for comparison. Column 1 (a-c): Increase in the dopant-free region from $W_i = 5$ nm to $W_i = 10$ nm (blue solid lines). Column 2 (d-f): Increase in the barrier height from $V_B = 0.15$ eV to $V_B = 0.25$ eV (red-solid lines). Column 3 (g-i): Increase in the ratio of the thermal conductivities of the well and barrier from $\kappa_W/\kappa_B = 5$ to $\kappa_W/\kappa_B = 10$ (green-solid lines). Column 4 (j-l): Increase in all the three parameters simultaneously to the levels of the data in Columns 1-3 (magenta-solid lines).

Figure 6:

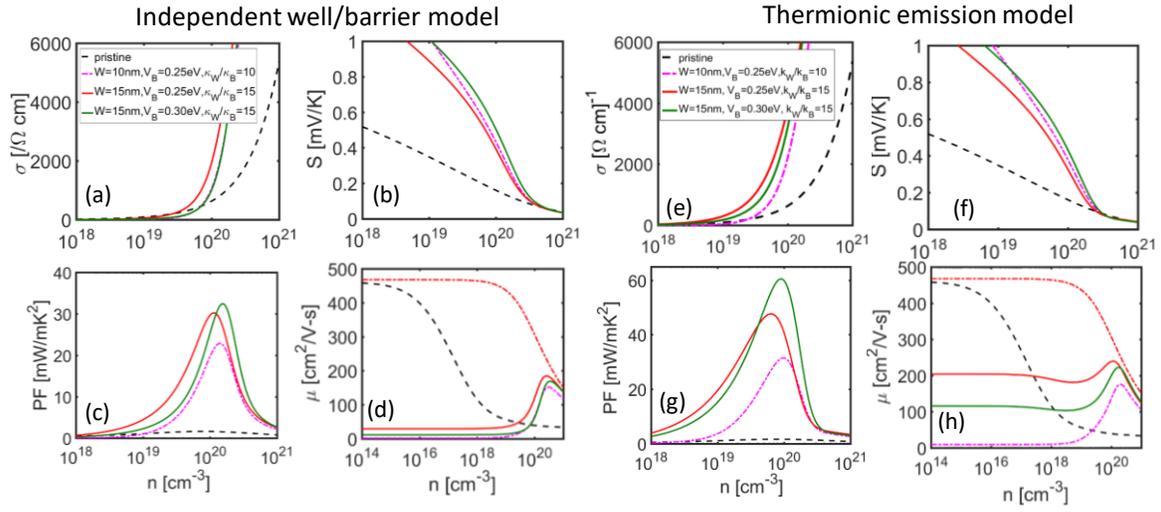

## Figure 6 caption:

The proposed design structure 'ingredients' increased one-by-one column-wise to extreme values of $W_i$ = 15 nm, $V_B$ = 30 eV, and $\kappa_W/\kappa_B$ = 15. (a-d) Undoped barrier, independent well/barrier model: (a) The electrical conductivity. (b) The Seebeck coefficients. (c) The power factors, (d) The mobilities versus carrier density. In all cases, the black-dashed lines shows the properties of the pristine flat band channel, without any barriers included and the magenta-dashed lines are the maximum power factor data from Fig. 5 for comparison. (e-h) The same quantities for the thermionic emission model.

Figure 7:

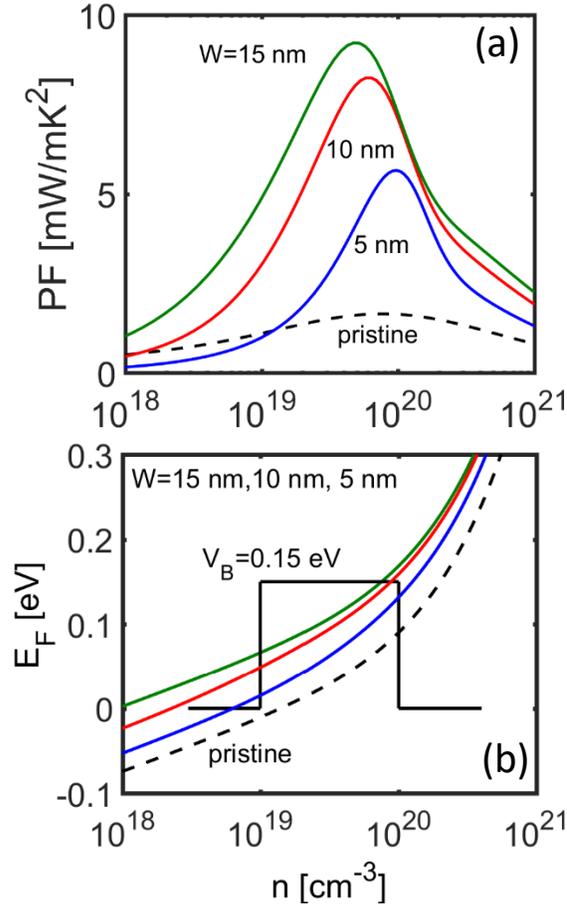

Figure 7 caption:

(a) The power factors of the structure with $V_B = 0.15$ eV in the case of dopant depleted regions of length $W_i$ = 5 nm, 10 nm, and 15 nm (the blue-solid line is repeated from Fig. 4i). (b) The position of the Fermi level $E_F$ from the bottom of the well in each case, versus carrier density. In both sub-figures the black-dashed line shows the properties of the pristine flat band channel, without any barriers included for comparison. The presence of the dopant free region, forces the $E_F$ higher compared to the pristine channel to achieve the same carrier density level.

Figure 8:

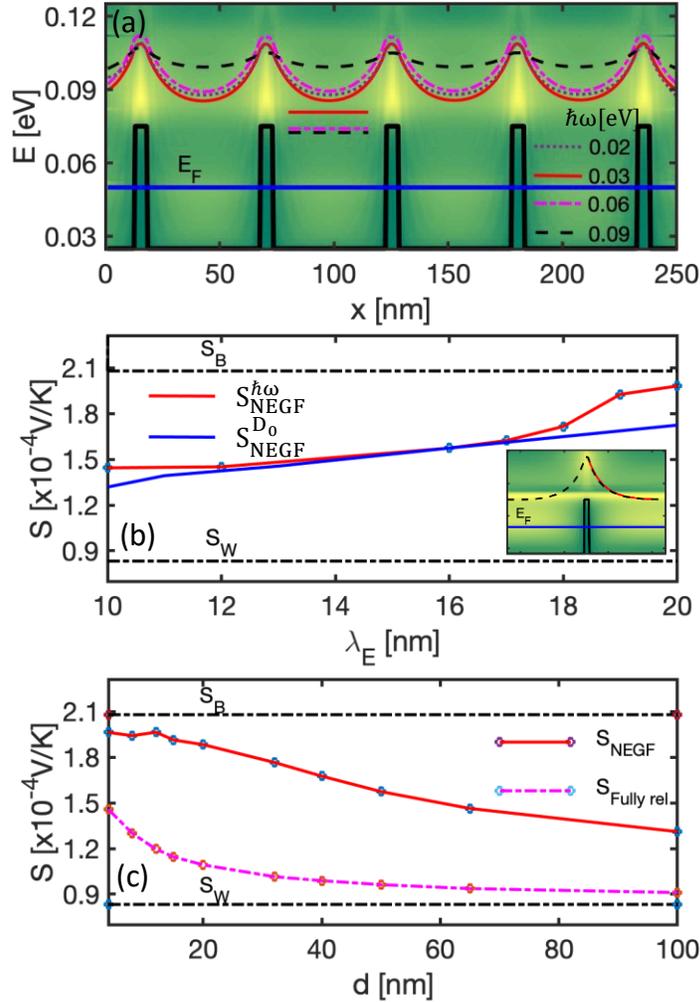

Figure 8 caption:

(a) The superlattice structure with the energy of the current flow <E> for different phonon-energies 0.02 eV, 0.03 eV, 0.06 eV, and 0.09 eV. The short horizontal dashed lines indicate the energy of the current level in the pristine channel, i.e. where <E> would relax to in an infinitely long well. The colormap indicates the current flow I(E,x). (b) The Seebeck coefficient versus the energy relaxation length of a superlattice for the case where $\lambda_E$ is altered by changing the phonon-energies (red line), and by altering the electron-phonon coupling strength (blue line). Inset: The extraction of $\lambda_E$ by an exponential fit of <E(x)> after the current passes over a single barrier. (c) The Seebeck coefficient of the superlattice structure versus the length of the well $L_W$ (red line). The magenta line shows the Seebeck coefficient in the case of full and immediate relaxation after the carriers pass over the barriers. The dashed horizontal lines in (b) and (c) indicate the Seebeck coefficient of a pristine material without barriers $S_W$ (infinite well), and of a pristine material with a large barrier $S_B$ (infinite barrier).

Figure 9:

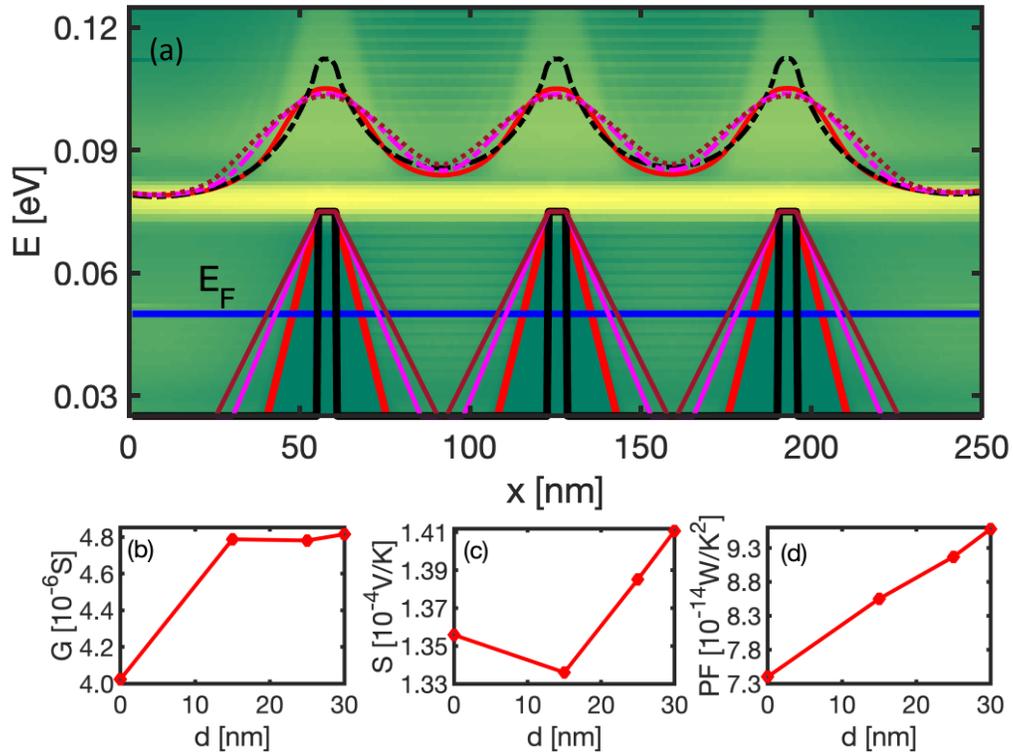

Figure 9 caption:

(a) The energy dependence of the current flow (yellow-green colormap) in superlattice (SL) structures with oblique barrier sidewalls, and the average energy of the current flow $<E(x)>$ (curved lines). The Fermi level is depicted by the flat blue line. The coloring of $<E>$ corresponds to the coloring of the barriers. (b-d) The conductance, Seebeck coefficient, and power factor of the SLs as a function of the sidewall inclination distance, *d*.

Figure 10:

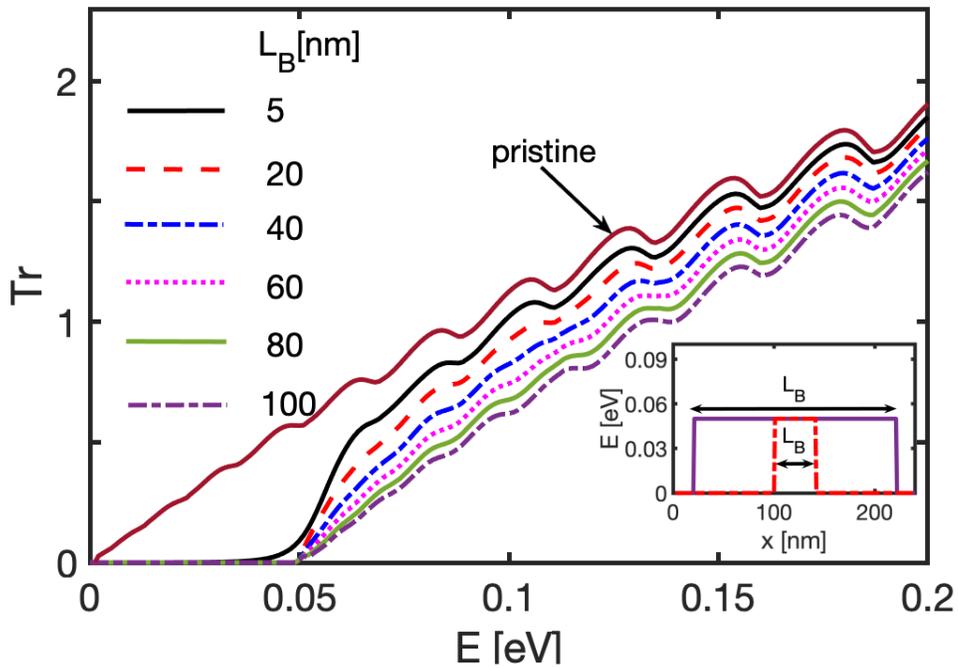

Figure 10 caption:

Indication of the degree of relaxation of carriers on the barrier material as they propagate over it. The figure shows the NEGF calculated energy resolved transmission function $Tr$ of the carriers in a channel with a single potential barrier with length $L_B$ as indicated in the inset. Electron scattering with acoustic phonons only are considered in the calculation. Cases for different barrier lengths are shown from a large $L_B$ = 100 nm taking over most of the channel (black-dashed line), to a pristine channel (brown line). The 'jump' of the $Tr$ from that of the larger barrier to that of the pristine material would indicate that carriers are more thermionically emitted over the barrier rather than relaxing on it as the barrier length is scaled.

Figure 11:

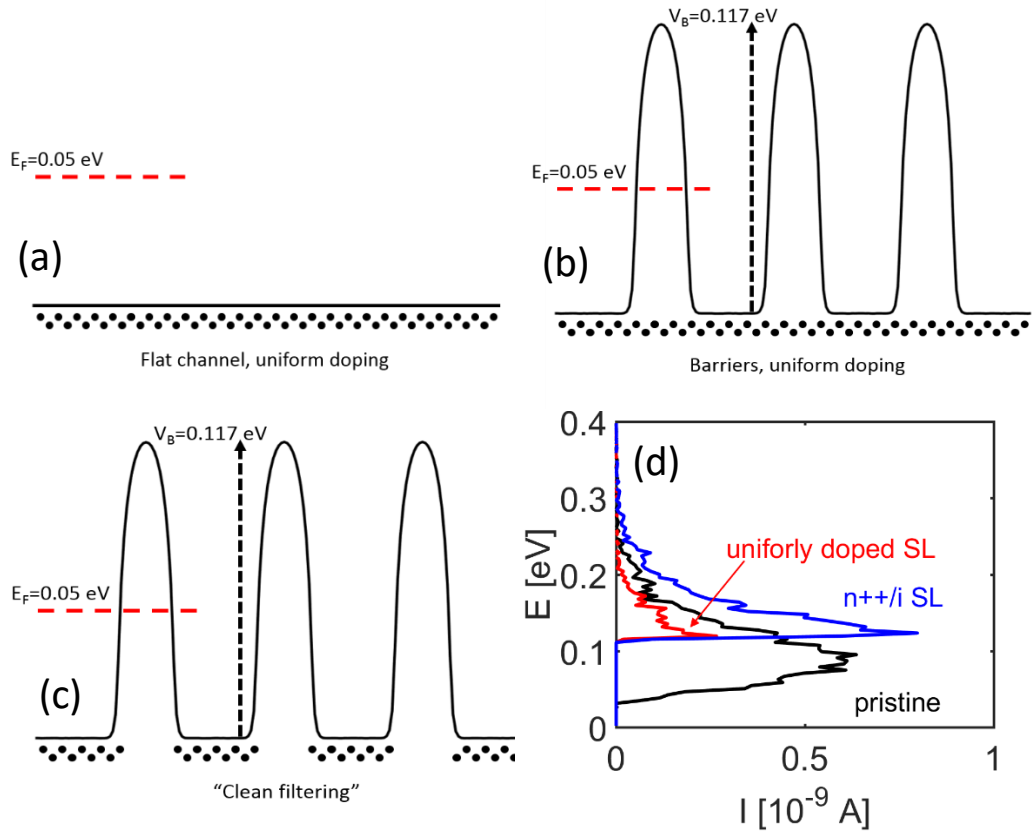

Figure 11 caption:

Schematics of structures simulated within Monte Carlo. (a) Pristine, uniformly doped channel (the dots indicate the doping placement). (b) Uniformly doped superlattice material. (c) Superlattice material consisting of a series doped/intrinsic regions (n++/i). The position of the Fermi level is indicated. The barrier shape is extracted from self-consistent solution of the Poisson equation (and used in (b) as well). (d) The energy resolved current in the three channels, in black for the pristine channel, in red for the uniformly doped SL, and in blue for the n++/i superlattice.